\documentclass[preprint,aps,floats,nofootinbib,amssymb,amsmath]{revtex4}
\usepackage{epsf,epsfig}

\newcommand{\be}{\begin{equation}}
\newcommand{\ee}{\end{equation}}
\newcommand{\bea}{\begin{eqnarray}}
\newcommand{\eea}{\end{eqnarray}}
\newcommand{\nn}{\nonumber}
\newcommand{\gev}{~\rm GeV}
\newcommand{\phix}{\phi_X}
\newcommand{\psix}{\psi_X}
\newcommand{\up}{\boldsymbol{U_p}}
\newcommand{\rp}{\boldsymbol{R_p}}
\newcommand{\ztwo}{\boldsymbol{Z_2}}
\newcommand{\til}{\widetilde}

\preprint{
\hbox to \hsize{
\hfill$\vcenter{
        \hbox{\bf UFIFT-HEP-07-15}
        \hbox{October 2007}}$}
}

\begin{document}

\title{
\vspace*{.75in}
A Supersymmetric $\boldsymbol{U(1)'}$ Model with Multiple Dark Matters
}

\author{Taeil Hur$^1$}
\author{Hye-Sung Lee$^2$}
\author{Salah Nasri$^{2,3}$}
\affiliation{
$^1$Department of Physics, KAIST, Daejon 305-701, Korea \\
$^2$Institute for Fundamental Theory, University of
Florida, Gainesville, FL 32611, USA \\
$^3$Department of Physics, United Arab Emirates University, Al Ain, UAE
\vspace*{.5in}}
\thispagestyle{empty}

\begin{abstract}
\noindent
We consider a scenario where a supersymmetric model has multiple dark matter particles.
Adding a $U(1)'$ gauge symmetry is a well-motivated extension of the Minimal Supersymmetric Standard Model (MSSM).
It can cure the problems of the MSSM such as the $\mu$-problem or the proton decay problem with high-dimensional lepton number and baryon number violating operators which $R$-parity allows.
An extra parity ($U$-parity) may arise as a residual discrete symmetry after $U(1)'$ gauge symmetry is spontaneously broken.
The Lightest $U$-parity Particle (LUP) is stable under the new parity becoming a new dark matter candidate.
Up to three massive particles can be stable in the presence of the $R$-parity and the $U$-parity.
We numerically illustrate that multiple stable particles in our model can satisfy both constraints from the relic density and the direct detection,
thus providing a specific scenario where a supersymmetric model has well-motivated multiple dark matters consistent with experimental constraints.
The scenario provides new possibilities in the present and upcoming dark matter searches in the direct detection and collider experiments.
\end{abstract}

\maketitle


\section{Introduction}
\label{sec:introduction}
Current cosmological data \cite{Spergel:2006hy} indicate that about 23\% of the energy in the Universe is in the form of cold dark matter (CDM).
The origin and the nature of the CDM is one of the biggest puzzles in theoretical particle physics today.
Since all known particles are ruled out as dark matter candidates, the dark matter problem provides one of the strongest phenomenological motivations for new physics beyond the Standard Model (SM).
Among the many possibilities, supersymmetry (SUSY) is perhaps one of the best motivated new physics scenarios: it resolves the fine-tuning problem of the SM, and may provide natural dark matter candidates among the spectrum of new particles.

It is not SUSY itself that guarantees the stability of the lightest supersymmetric particle (LSP), but rather an additional $\ztwo$ symmetry ($R$-parity \cite{Rparity}) which is introduced to solve the proton decay problem: without $R$-parity or some similar type of stabilization mechanism, the superpartners of the SM fermions would be expected to mediate proton decay at an unacceptably high rate.
$R$-parity has attractive features.
First, it protects the proton from decaying via renormalizable lepton number violating operators and baryon number violating operators.
Second, it allows TeV scale SUSY to evade the stringent bounds from electroweak precision data.
Finally, in the presence of $R$-parity, the LSP is stable and can be a viable dark matter candidate.
For all these reasons, the Minimal Supersymmetric Standard Model (MSSM) with $R$-parity has been the most popular supersymmetric extension of the SM, and most SUSY dark matter studies have been confined to this setup \cite{Jungman:1995df}.
Within the MSSM, the lightest neutralino (a mixture of neutral gauginos and higgsinos) has been the only viable CDM candidate since the other possibility (the left-handed sneutrinos) has been ruled out by combining two constraints for a viable dark matter: relic density and direct detection \cite{Falk:1994es}.

In spite of its success and popularity, the MSSM with $R$-parity still has two major problems: the $\mu$-problem \cite{Kim:1983dt} and the potential proton decay problem due to dimension 5 operators \cite{dim5pdecay}.
The MSSM does not explain why the only mass parameter in the superpotential ($\mu$ parameter) is at the electroweak scale instead of the fundamental (Planck or GUT) scale.
Furthermore, $R$-parity allows for dimension 5 lepton and baryon number violating operators such as $QQQL$ and $U^cU^cD^cE^c$ which may still violate the bound on the proton lifetime.
A natural solution to these problems would probably require that the MSSM be extended by a new mechanism or a new symmetry.
The $U(1)'$-extended MSSM (UMSSM) \cite{UMSSM} is a straightforward extension of the MSSM with a non-anomalous TeV scale Abelian gauge symmetry.
It can solve the $\mu$-problem and the dimension 5 operator problem very naturally with an appropriate charge assignment.
An extra Abelian gauge symmetry is also suggested by many new physics scenarios including grand unification \cite{GUT_U1}, extra dimension \cite{Masip:1999mk}, superstring \cite{Cvetic:1995rj}, little Higgs theories \cite{littleHiggs}, strong dynamics \cite{Hill:2002ap}, and Stueckelberg mechanism \cite{Stueckelberg}.

The UMSSM retains most attractive features of the MSSM, and also brings new aspects in relation to the dark matter problem.
First, it extends the set of possible LSP dark matter candidates: the superpartners of the Higgs singlet $S$ and the $Z'$ gauge boson are now additional components of the neutralinos \cite{UMSSMneutralino}.
The (predominantly right-handed) sneutrino also becomes a thermal CDM candidate if it couples to $Z'$ \cite{Lee:2007mt}.
Furthermore, a new gauge symmetry also naturally introduces exotic particles required to cancel gauge anomalies \cite{exotics}.
When a neutral massive field is among the exotics, it could be a CDM candidate as well: it could be the LSP and stable due to $R$-parity, but it is also plausible that it is not the LSP, yet still is stable due to a new discrete gauge symmetry \cite{IbanezRoss, Luhn}
, which may emerge naturally as a residual symmetry after the $U(1)'$ gauge symmetry is broken.
Because of its gauge origin, the discrete gauge symmetry does not suffer from domain wall problem \cite{domainwall} or symmetry violation by gravity effects \cite{Krauss:1988zc}.
With the presence of two discrete symmetries very naturally introduced, the dark matter sector is enriched with coexisting multiple dark matter particles.
Most previous studies on SUSY dark matter assume that the dark matter is made up of a single component.
Then it is usually required that the LSP makes up 100\% of the CDM in the Universe or, if the nature of the LSP is such that its annihilation rate is too high, SUSY dark matter is simply treated as a subdominant component of the CDM, without discussing the dominant component or providing the quantitative consistency with experimental constraints.

In this paper, we consider the $U(1)'$-extended MSSM with a residual discrete gauge symmetry ($\ztwo$ for simplicity) arising from the $U(1)'$ gauge symmetry.
We consider a simple case where a SM singlet exotic is a stable particle under the new $\ztwo$ symmetry.
Together with the LSP which is stable under $R$-parity, there can be up to three dark matters due to the kinematics (see Appendix \ref{sec:triangle}).
We also illustrate explicit examples for multiple dark matters that satisfy the relic density and direct detection constraints forming a viable texture of dark matters in the Universe.

In Section \ref{sec:UMSSM}, we describe the possible remnant discrete symmetry of the $U(1)'$ extension of the MSSM.
In Section \ref{sec:relic}, we discuss the relic density and annihilation channels.
In Section \ref{sec:direct}, we discuss the direct detection.
In Section \ref{sec:analysis}, we perform numerical analysis on the relic density and the direct detection.
In Section \ref{sec:conclusion}, we summarize our results.
In Appendix \ref{sec:triangle}, we list the conditions where the multiple particles are stable.
In Appendix \ref{sec:lagrangians}, we list the relevant lagrangians in our model.

\section{$U$-parity and the lightest $U$-parity particle}
\label{sec:UMSSM}
There are already at least two SM singlet chiral superfields in our model, the Higgs singlet $S$ that spontaneously breaks the new symmetry with its vacuum expectation value (vev) and the right-handed neutrino $N^c$ that explains the neutrino mass.
Higgs is not suitable for the odd particle under a parity since its replacement with a vev would destroy the discrete symmetry.
Right-handed neutrino that forms $LH_2N^c$ is not good either for the odd particle since lepton doublet also should have the odd parity and it is the active light neutrino that would be the lightest odd particle under new parity.
We will consider a new SM singlet ($X$) -- neither $S$ nor $N^c$ -- that may be required from the anomaly cancellation condition, as our new dark matter candidate\footnote{For an example of a SM singlet required for the anomaly cancellation in a $U(1)'$-extended MSSM, see Appendix III of Ref.~\cite{Lee:2007fw}.}.

It is natural to think about possible remnant discrete symmetry \cite{IbanezRoss, Luhn} from our $U(1)'$ gauge symmetry.
A discrete gauge symmetry $\boldsymbol{Z_N}$ would emerge if the discrete charges and the $U(1)'$ charges satisfy the following relation:
\be
z[S] = N, \qquad
z[F_i] = q[F_i] + n_i N
\label{eqn:chargerelation}
\ee
where $z[F_i]$ and $q[F_i]$ stand for the $U(1)'$ charge and the $\boldsymbol{Z_N}$ charge for a field $F_i$, respectively.
$z[F_i]$, $q[F_i]$, $n_i$ and $N$ are all integers after an appropriate normalization of charges. 
Higgs singlet $S$ is supposed to have $q[S] = 0~ (\text{mod}~ N)$ to keep the discrete symmetry after it is replaced by its vev (e.g. both $SXX$ and $\left<S\right>XX$ should be singlet under the discrete symmetry).

We assume a $\ztwo$ ($N=2$) discrete parity, which we shall call $U$-parity ($\up$), as a remnant discrete symmetry of the $U(1)'$ gauge symmetry.
We assign even $\up$ to the SM fields, and odd $\up$ to the $X$ field.
We assume either a scalar ($\phix$) or a fermion ($\psix$) component (whichever lighter) of the superfield $X$ is the Lightest $U$-parity Particle (LUP).
The discrete charges are then
\be
q[X] = 1, \qquad
q[{\rm SM~ fields}] = 0 \qquad (\text{mod}~ 2) \ . \label{eqn:discretecharges}
\ee
The $\ztwo$ symmetry allows terms that contains $X$ only in even numbers such as $SXX$.
To satisfy the condition~(\ref{eqn:chargerelation}) of the discrete gauge symmetry, we need to assign $U(1)'$ charges as $z[S]$ multiplied by an integer to the SM fields, but not to the $X$ field.
We assume $SXX$ is the only term with $X$ in the superpotential (thus, $\psix$ is a Majorana particle), and we can assign either even or odd $R$-parity ($\rp$) to $\phix$ (and the other to $\psix$).
It is obvious that the MSSM and the UMSSM superpotentials can allow such a charge assignment.

To be an anomaly-free theory, the $U(1)'$ charges should satisfy the gauge anomaly conditions when whole particle spectrum is considered.
We do not restrict ourselves to a specific particle spectrum, and just require the LUP to be a SM singlet exotic, while various exotic particles may exist with odd or even $U$-parity.
We assume all exotics are vector-like for the SM gauge group so that the SM gauge group  anomaly conditions are not altered.
If a $\boldsymbol{Z_N}$ is a discrete gauge symmetry originated from a $U(1)'$ gauge symmetry, the discrete charges should satisfy the discrete anomaly conditions \cite{IbanezRoss, Luhn}:
\bea
\left[SU(3)_C\right]^2 - U(1)' &:& \sum_{i=\text{quark}} q_i = 0 \quad (\text{mod}~ N) \\
\left[SU(2)_L\right]^2 - U(1)' &:& \sum_{i=\text{doublet}} q_i = 0 \quad (\text{mod}~ N) \\
\left[\text{gravity}\right]^2 - U(1)' &:& \sum_i q_i = 0 \quad (\text{mod}~ N) \\
\left[U(1)_Y\right]^2 - U(1)' &:& \sum_i y_i^2 q_i = 0 \quad (\text{mod}~ N) \\
U(1)_Y-\left[U(1)'\right]^2 &:& \sum_i y_i q_i^2 = 0 \quad (\text{mod}~ N) \\
\left[U(1)'\right]^3 &:& \sum_i q_i^3 = 0 \quad (\text{mod}~ N)
\eea
where $i$ runs through all SM fields and the exotic fields, and $y_i$ stand for the hypercharges normalized to integers for all fields.
If there is only one exotic field $X$, the discrete charges of eq.~(\ref{eqn:discretecharges}) satisfy theses discrete anomaly conditions automatically except for the $\left[\text{gravity}\right]^2$-$U(1)'$ and $\left[U(1)'\right]^3$, which means we need more exotics with nonzero discrete charges.
We will not give further constraints from the full gauge anomaly-free conditions keeping the possibility of various exotic fields open.

Although $S H_2 H_1$ term (i.e. effective $\mu$ term) greatly motivates the low energy scale of the $U(1)'$ resolving the $\mu$-problem, it is technically involved to perform numerical analysis for all channels when the Higgs doublets and the Higgs singlet are coupled (through Yukawa term as well as $D$-term).
Therefore, for the sake of simplicity, we will limit our numerical analysis to the case of $z[H_1] = z[H_2] = 0$ with which the $\mu$-problem is not solved in the way of Ref.~\cite{UMSSM}, i.e. by forbidding the original $\mu$ term $H_2 H_1$ and allowing the effective $\mu$ term $S H_2 H_1$.
The analysis of the UMSSM and the LUP dark matter in a more general setup where the Higgs doublets have non-vanishing $U(1)'$ charges and $\mu$-problem is resolved will be discussed in other publication.
In the limit of vanishing charges for Higgs doublets, $S H_2 H_1$ term as well as $S S^* H_i H_i^*$ terms from the $D$-term will be forbidden making the Higgs singlet $S$ an isolated physical eigenstate and Higgs doublets $H_1$, $H_2$ the same as those of the MSSM. (For general properties of Higgses in the UMSSM in comparison with other models, see Ref.~\cite{UMSSMHiggs}.)
The $Z$ and $Z'$ mixing will be zero, and the $Z'$-ino ($\til Z'$) and singlino ($\til S$) components of the neutralino sector are also completely decoupled from the MSSM neutralinos since the mixing terms are proportional to $z[H_i]$.

The superpotential is then given by, assuming three right-handed neutrino Dirac mass terms,
\be
W = \mu H_2 H_1 + y_U H_2 Q U^c + y_D H_1 Q D^c + y_N H_2 L N^c + y_E H_1 L E^c + \frac{k}{2} S X X \ .
\ee
Here, the Yukawa terms give the following relations among family universal\footnote{Family non-universal $U(1)'$ charges may induce dangerous flavor-changing neutral currents \cite{Langacker:2000ju}. However, the flavor-changing $Z'$ may explain the discrepancies in rare $B$ decays with appropriate parameter choices \cite{Banomaly}.} $U(1)'$ charges:
\bea
\label{eqn:charge_assign}
&&z[Q] = -z[U^c] = -z[D^c] = n_Q z[S] \\
&&z[L] = -z[N^c] = -z[E^c] = n_L z[S] \\
&&z[X] = -\frac{1}{2} z[S]
\eea
where $n_Q$ and $n_L$ are integers.
With these charge assignments, Majorana neutrino mass term $N^c N^c$ (unless $n_L = 0$) or $S N^c N^c$ are not allowed in general.
We will assume neutrinos are Dirac throughout this paper\footnote{Additional relativistic degrees of freedom would contribute to the $^4$He abundance, but be diluted by large $Z'$ mass. It may explain the discrepancy of the $^4$He measurement \cite{superweakBBN, Barger:2003zh}.}.
Therefore, we have three free parameters (including $z[S]$ not normalized to $N$) for the $U(1)'$ charges in our numerical study: $n_Q$, $n_L$, $z[S]$.

The dangerous dimension 5 proton decay operators $QQQL$, $U^cU^cD^cE^c$, and $U^cD^cD^cN^c$ have commonly total $U(1)'$ charge $(3 n_Q + n_L) z[S]$ (up to overall sign) with the above charge assignment, and  are forbidden if condition 
\be
3 n_Q + n_L \ne 0  \label{eqn:dim5}
\ee
is satisfied.

In the $R$-parity violating supersymmetric model or any other model which does not have a stable dark matter candidate, the LUP can provide a good CDM candidate with an addition of the $U(1)'$ gauge symmetry.
Depending on the detail of charges and spectrum, the $U(1)'$ may also ensure longevity of the proton without $R$-parity (for example, see Ref.~\cite{Lee:2007fw}).

\section{Relic density}
\label{sec:relic}
The relic density of dark matter is precisely measured as $\Omega_\text{CDM} h^2 = 0.111^{+0.011}_{-0.015}$ ($2\sigma$ allowed range) by the WMAP+SDSS \cite{Spergel:2006hy}.
Present day relic density of a dark matter component is given by
\be
\Omega h^2 = \frac{8 \pi}{3} \frac{s(T_0) M}{M_{Pl}^2 ( 100~ \text{km/s/Mpc})^2} Y(T_0)
             = 2.742 \times 10^8 \frac{M}{\gev} Y(T_0)
\ee
where $M_{Pl}$ is Plank mass, $s(T_0)$ is the entropy density at present time, $h$ is the normalized Hubble constant and the relic abundance $Y(T)$ is defined as the number density divided by the entropy density. The abundance of the dark matter $Y(T)$ can be calculated by solving the Boltzmann equation
\be
\frac{dY}{dT} = \sqrt{\frac{\pi g_*(T)}{45}} M_{Pl} \left<\sigma v \right> (Y(T)^2 - Y_{\text{eq}}(T)^2)
\ee
where $g_*$ is an effective number of relativistic degree of freedom and $Y_{\text{eq}}(T)$ is the thermal equilibrium abundance. 
$\left<\sigma v \right>$ is the thermally averaged annihilation cross-section times relative velocity.

For our numerical calculation, we use {\tt micrOMEGAs} \cite{Belanger:2006is} which performs rapid relic density calculation for $2 \to 2$ processes.
We implement, on top of the MSSM fields and interactions, additional fields and interactions of the UMSSM as well as three right-handed Dirac neutrinos with negligible masses.
The relevant lagrangians of the model are listed in Appendix \ref{sec:lagrangians}.

We take vanishing soft trilinear terms limit ($A = 0$) in our numerical analysis.
A non-vanishing $A_{SXX} S\phix\phix$ soft term would separate masses of two components of complex $\phix$ field, after $S$ is replaced by its vev, which we want to avoid for numerical simplicity.
With assumption of the zero charges for Higgs doublets (as discussed in Section \ref{sec:UMSSM}), and sufficient mass splittings (so that co-annihilations are irrelevant), the annihilation channels for $\psix$ and $\phix$ are given as follows.
\begin{enumerate}
\item $\psix \psix \to f \bar f$ ($Z'$ mediated $s$-channel)
\item $\psix \psix \to \til f \til f^*$ ($S$ mediated $s$-channel, $Z'$ mediated $s$-channel)
\item $\psix \psix \to S S$, $Z' Z'$ ($S$ mediated $s$-channel, $\psix$ mediated $t$-channel)
\item $\psix \psix \to S Z'$ ($Z'$ mediated $s$-channel, $\psix$ mediated $t$-channel)
\item  $\psix \psix \to \til S \til S$ ($Z'$ mediated $s$-channel, $\phix$ mediated $t$-channel)
\item  $\psix \psix \to \til Z' \til Z'$ ($\phix$ mediated $t$-channel)
\item  $\psix \psix \to \til S \til Z'$ ($S$ mediated $s$-channel, $\phix$ mediated $t$-channel)
 \item  $\psix \psix \to \phix \phix^*$ ($S$ and $Z'$ mediated $s$-channel, $\til S$ and $\til Z'$ mediated $t$-channel)
\item  $\psix \psix \to \phix \phix(\phix^* \phix^*)$ ($\til S$ and $\til Z'$ mediated $t$-channel)
\item $\phix \phix^* \to f \bar f$ ($Z'$ mediated $s$-channel)
\item $\phix \phix^* \to \til f \til f^*$ ($S$ mediated $s$-channel, $Z'$ mediated $s$-channel, 4 point interaction)
\item $\phix \phix^* \to S S$ ($S$ mediated $s$-channel, $\phix$ mediated $t$-channel, 4 point interaction)
\item $\phix \phix^* \to Z'Z'$ ($S$ mediated $s$-channel, $\phix$ mediated $t$-channel, 4 point interaction)
\item $\phix \phix^* \to S Z'$ ($Z'$ mediated $s$-channel, $\phix$ mediated $t$-channel)
\item $\phix \phix^* \to \til S \til S$ ($Z'$ mediated $s$-channel, $\psix$ mediated $t$-channel)
\item $\phix \phix^* \to \til Z' \til Z'$ ($\psix$ mediated $t$-channel)
\item $\phix \phix^* \to \til S \til Z'$ ($S$ mediated $s$-channel, $\psix$ mediated $t$-channel)
\item $\phix \phix^* \to \psix \psix$ ($S$ and $Z'$ mediated $s$-channel, $\til S$ and $\til Z'$ mediated $t$-channel)
\item  $\phix \phix(\phix^* \phix^*) \to \psix \psix$ ($\til S$ and $\til Z'$ mediated $t$-channel)
\item  $\phix \phix(\phix^* \phix^*) \to \til S \til S$, $\til Z' \til Z'$, $\til S \til Z'$ ($\psix$ mediated $t$-channel)
\end{enumerate}
The $\psix$ annihilation includes features of Ref.~\cite{Brahm:1989jh}, and $\phix$ annihilation includes features of Ref.~\cite{Lee:2007mt}.
Additional neutralinos $\til Z'$ and $\til S$ mix with each other, and we will call the mass eigenstates of them as $\chi'$ from now on.

\section{Direct detection}
\label{sec:direct} 
There are many direct detection experiments that attempt to detect the dark matter particles via nuclear recoil as the Earth passes through dark matter halo of our galaxy (for examples, see Refs. \cite{Bernabei:2000qi, Sanglard:2005we, Akerib:2005kh, Lee.:2007qn, Angle:2007uj}).
Most experimental limits of the direct detection are given in terms of the cross-section per nucleon on the assumption of sole dark matter.
Here, we will first discuss the general formalism for multiple dark matters.

Let us introduce a parameter $\epsilon_i$ which parametrizes the fraction of the energy density of a dark matter $i$ in our local dark matter halo and also in the whole universe.
\begin{equation}
\epsilon_i = \frac{\rho_i^{\text{halo}}}{\rho_{\text{CDM}}^{\text{halo}}} \simeq
\frac{\Omega_i}{\Omega_{\text{CDM}}}
\end{equation}
where $\Omega_{\text{CDM}} h^2 = \sum_i {\Omega_i h^2}$, and  $\sum_i {\epsilon_i} = 1$.
The event rate per unit detector mass per unit time is given, from a dimensional analysis, by
\be
R \sim n \sigma \left<v\right> / M_N
\ee
where $M_N$ is a nucleus mass and $n$ is a number density of the dark matter.
A precise differential rate \cite{Jungman:1995df}, for each component $i$, is given by
\be
\frac{dR_i}{dE_R} = \frac{\sigma_{0, i} \rho_i}{\sqrt{\pi} v_0 M_i \mu_{N, i}^2} F^2(E_R) T_i(E_R)
\ee
where $\sigma_{0}$ is the elastic scattering cross-section of a dark matter off a nucleus, $\mu_{N} = M M_N / (M+M_N)$ is the effective mass of a dark matter and a nucleus, and $v_0 \simeq 220 ~\text{km/s}$ is the circular speed of the Sun around the galactic center (neglecting the relative velocity of the Sun and the Earth).

Let us assume a simple form factor of $F(E_R) \approx 1$, and a pure Maxwellian speed distribution which gives
\be
T_i(E_R) = \exp{(-v_{i, \text{min}}^2/v_0^2)}
\ee
where $v_{i, \text{min}} =  \sqrt{\frac{E_T M_N}{2\mu_{N, i}^2}}$ is the minimum velocity of a CDM that can produce an energy deposit above the detector threshold energy $E_T$. 
Then the direct detection rate for each component is
\be
R_i =  \int_{E_T}^{\infty} \frac{dR_i}{dE_R} dE_R = \frac{\sigma_{0, i} \rho_i}{M_i M_N}\frac{2v_0}{\sqrt{\pi}}\exp{(-E_T/E_0^i)}
\ee
where $E_0 = 2\mu_{N}^2v_0^2/M_N$, and the total detection rate is a sum of the contributions from each component.
This total rate is the quantity constrained by the experiments.
\be
R = \sum_i R_i = C\sum_i{\frac{\epsilon_i\sigma_{0, i}}{M_i}\exp{(-E_T/E_0^i)}} < R_{\text{exp}}
\ee
where $C = 2\rho_{\text{CDM}}^{\text{halo}}v_0/(\sqrt{\pi} M_N)$.

Because of different masses of the CDM species, the usual constraint on the cross-section is not applicable in general for multiple dark matter cases.
However, in some special cases where the relevant mass is effectively only one, those constraints can still be used.
For example, let us consider a special case of where only one dark matter component $(\xi)$ interacts with a nucleus while the others do not.
The experimental bound on the detection rate $R < R_{\text{exp}}$ can be rewritten for the cross-section
\be
\epsilon \sigma_{0} < M \exp{(E_T/E_0)} C^{-1}R_{\text{exp}}
\label{eqn:effsigma}
\ee
where $\epsilon$, $\sigma_{0}$, $E_0$, $M$ are those of the dark matter $\xi$.
It is the cross-section per nucleon that is usually compared to the experimental data, which is obtained by
\be
\sigma_n = \frac{\mu_n^2}{\mu_N^2} \frac{\sigma_0}{A^2}
\ee
where $\mu_n = M M_n/ (M + M_n)$.
The {\em effective cross-section} for the single detectable dark matter case is then given by $\sigma_{n, \text{eff}} = \epsilon\sigma_{n}$, and the experimental constraint is imposed on $\epsilon \sigma_{n}$ instead of $\sigma_{n}$.

Returning to our model, the dominant channels for the direct detection of $\psix$ and $\phix$ are $Z'$ mediated $t$-channels.
If we had $SH_2H_1$ term, there would have been other channels such as $H_i$ mediated $t$-channel (where $H_i$ is a general mixture of Higgs doublets and a Higgs singlet).
For the MSSM LSP, we consider the lightest neutralino ($\chi^0$).

The effective low-energy lagrangian for elastic scattering of the CDM and a quark is given by
\be
{\cal{L}}_{\text{eff}} = {\cal{L}}_{\psix-q} + {\cal{L}}_{\phix-q} + {\cal{L}}_{\chi^0-q}
\ee
where
\bea
{\cal{L}}_{\psix-q} &=&
 G_{Z'}z[X]z_A[q]\overline{\psix}\gamma^{\mu}\gamma_5\psix\overline{q}\gamma_{\mu}\gamma_5q\\
{\cal{L}}_{\phix-q} &=&
 iG_{Z'}z[X]\left(\phix^*\partial_{\mu}\phix - \partial_{\mu}\phix^*\phix \right) \left({z_V[q]\overline{q}\gamma^{\mu}q} + {z_A[q] \overline{q}\gamma^{\mu}\gamma_5q}\right)\\
{\cal{L}}_{\chi^0-q} &=& f_{q}\overline{\chi^0}\chi^0\overline{q}q + d_{q}\overline{\chi^0}\gamma_{\mu}\gamma_5\chi^0\overline{q}\gamma^{\mu}\gamma_5q
\eea
and $G_{Z'} = g^2_{Z'}/{M^2_{Z'}}$, $z_V[q] = \frac{1}{2} (z[q_L] + z[q_R])$, $z_A[q] = \frac{1}{2} (z[q_R] - z[q_L])$. 
The $f_q$ term results from Higgs exchange and squark exchange.
The $d_q$ term is due to $Z$ exchange and squark exchange. 
In the non-relativistic limit the $\psix$ interaction and the $d_{q}$ term are spin-dependent interactions, while the other interactions are spin-independent and dominate due to the coherence effect.

With a choice of the MSSM-like lightest neutralino (i.e. mass eigenstates of $\til S$ and $\til Z'$ are heavier than the lightest mass eigenstate of $\til B$, $\til W_3$, $\til H_1$, $\til H_2$) and heavy squarks, the effective scalar coupling of a the neutralino dark matter to up-type and down-type quarks are approximately given by \cite{Jungman:1995df}
\be
f_u \simeq \sum_{H_i = h, H} \frac{g_2 T_{H_i \chi^0 \chi^0} T_{H_i u u}}{2 M_{H_i}^2}, \qquad f_d \simeq \sum_{H_i = h, H} \frac{g_2 T_{H_i \chi^0 \chi^0} T_{H_i d d}}{2 M_{H_i}^2}
\ee
with
\bea
T_{h \chi^0 \chi^0} = (N_{40} \cos\alpha + N_{30} \sin\alpha) (N_{20} - \tan\theta_W N_{10}) \\
T_{H \chi^0 \chi^0} = (N_{40} \sin\alpha - N_{30} \cos\alpha) (N_{20} - \tan\theta_W N_{10})
\eea
\bea
T_{h u u} = -\frac{g_2 m_u \cos\alpha}{2 M_W \sin\beta} \qquad T_{h d d} = +\frac{g_2 m_d \sin\alpha}{2 M_W \cos\beta} \\
T_{H u u} = -\frac{g_2 m_u \sin\alpha}{2 M_W \sin\beta} \qquad T_{H d d} = -\frac{g_2 m_d \cos\alpha}{2 M_W \cos\beta}
\eea
where $\alpha$ is the mixing angle that diagonalizes the CP even Higgs mass matrix, and $\tan\beta = \left<H_2\right> / \left<H_1\right>$.

In the non-relativistic limit, the spin-independent cross-section of a CDM and target nucleus is given by
\be
\sigma_0^{SI} = \frac{\mu_N^2}{\pi} \left( Z \lambda_p + (A-Z) \lambda_n \right)^2
\ee
where
\bea
\psix: &&\lambda_p (\psix) = 0, \quad \lambda_n (\psix) = 0 \label{eq:psixsigma} \\
\phix: &&\lambda_p (\phix) = G_{Z'} z[X] ( 2 z_V[u] + z_V[d]), \quad \lambda_n (\phix) = G_{Z'} z[X] ( z_V[u] + 2 z_V[d]) \\
\chi^0: &&\lambda_p (\chi^0) = 2 f_p = 2 m_p \left( f_{T_s} \frac{f_s}{m_s} + \frac{2}{27} f_{TG} \sum_{q = c, b, t} \frac{f_q}{m_q} \right), \quad \lambda_n (\chi^0) = 2 f_n \simeq 2 f_p
\eea
with non-negligible nucleon parameters given by $f_{T_s} = 0.12$, $f_{TG} = 0.84$ ($s$-quark, heavy quarks contribution for both proton and neutron) \cite{Ellis:2000ds}.
The spin-independent cross-section for $\psix$ dark matter is zero.
Then, with all three dark matters present, the experimental bound can be 
written as 
\bea
\frac{\epsilon_{\phix}\sigma_{0, \phix}^{SI}}{M_{\phix}}\exp{(-E_T / E_0^{\phix})} +
\frac{\epsilon_{\chi^0}\sigma_{0, \chi^0}^{SI}}{M_{\chi^0}}\exp{(-E_T / E_0^{\chi^0})} < C^{-1}R_{\text{exp}}.
\eea

\section{Numerical Analysis}
\label{sec:analysis}
In this section, we present the numerical results of the relic densities and direct detection cross-sections.
For definiteness of our numerical analysis, we make following choices for our parameter values.
\begin{enumerate}
\item We assume the $U(1)'$ gauge coupling constant and charge assignments as following:
\be
g_{Z'} = g_1 = \sqrt{\frac{5}{3}} g_Y
\ee
\be
\label{eqn:charge_assign2}
n_Q = 1 \qquad n_L = 5 \qquad z[S] = 0.3
\ee
which correspond to
\be
z[q_L] = z[q_R] = 0.3 \qquad z[\ell_L] = z[\ell_R] = 1.5 \qquad z[X] = -0.15 \ .
\ee
The above choice of $n_Q$ and $n_L$ satisfies the condition (\ref{eqn:dim5}) to avoid dangerous dimension 5 proton decay operators, and the relatively small $U(1)'$ coupling to quarks help avoiding the direct detection constraints for the $\phix$ dark matter.

\item We choose the dark matters so that the lightest neutralino ($\chi^0$) is always the LSP, and either the $\psix$ (with $\rp$ even) or the $\phix$ (with $\rp$ even) is the LUP out of several possibilities.

\item We choose the parameter values as
\be
M_2 = \mu \quad \ll \quad M_1
\ee
so that $\chi^0$ is wino or higgsino-like.
The wino or higgsino-like LSP has been considered not a good cold dark matter candidate since their annihilation rate is too high to satisfy the measured CDM relic density.
In the multiple dark matters scenario, they are actually preferable LSP dark matter candidates that can coexist with other dark matter.
Since $M_2$ and $\mu$ are the diagonal components of the chargino mass matrix, we usually get a light chargino ($\chi_1^\pm$) also, and the co-annihilation between $\chi^0$ and $\chi_1^\pm$ becomes important for the relic density calculation of $\chi^0$.
We need to scan $M_2$ for analysis, but for other gaugino masses we choose
\be
M_1 = 1000 \gev \qquad M_3=1000 \gev \qquad M_{1'}=100 \gev
\ee
and choose $M_A = 500 \gev$, $\tan\beta =40$ and $M_S = 800 \gev$.

\item We assume degenerate masses for squarks and sleptons and vanishing trilinear terms 
\be
M_\text{squarks} = 1100 \gev \qquad M_\text{sleptons} = 600 \gev \qquad A_\text{all} = 0 \ .
\ee
The large squark masses will make the squark contribution to the direct detection small.

\item We assume all exotics (except $X$) that might be required for the anomaly cancellation are very heavy, and neglect their contributions to the relic density.

\end{enumerate}

\begin{figure}[t]
\includegraphics[width=0.49\textwidth]{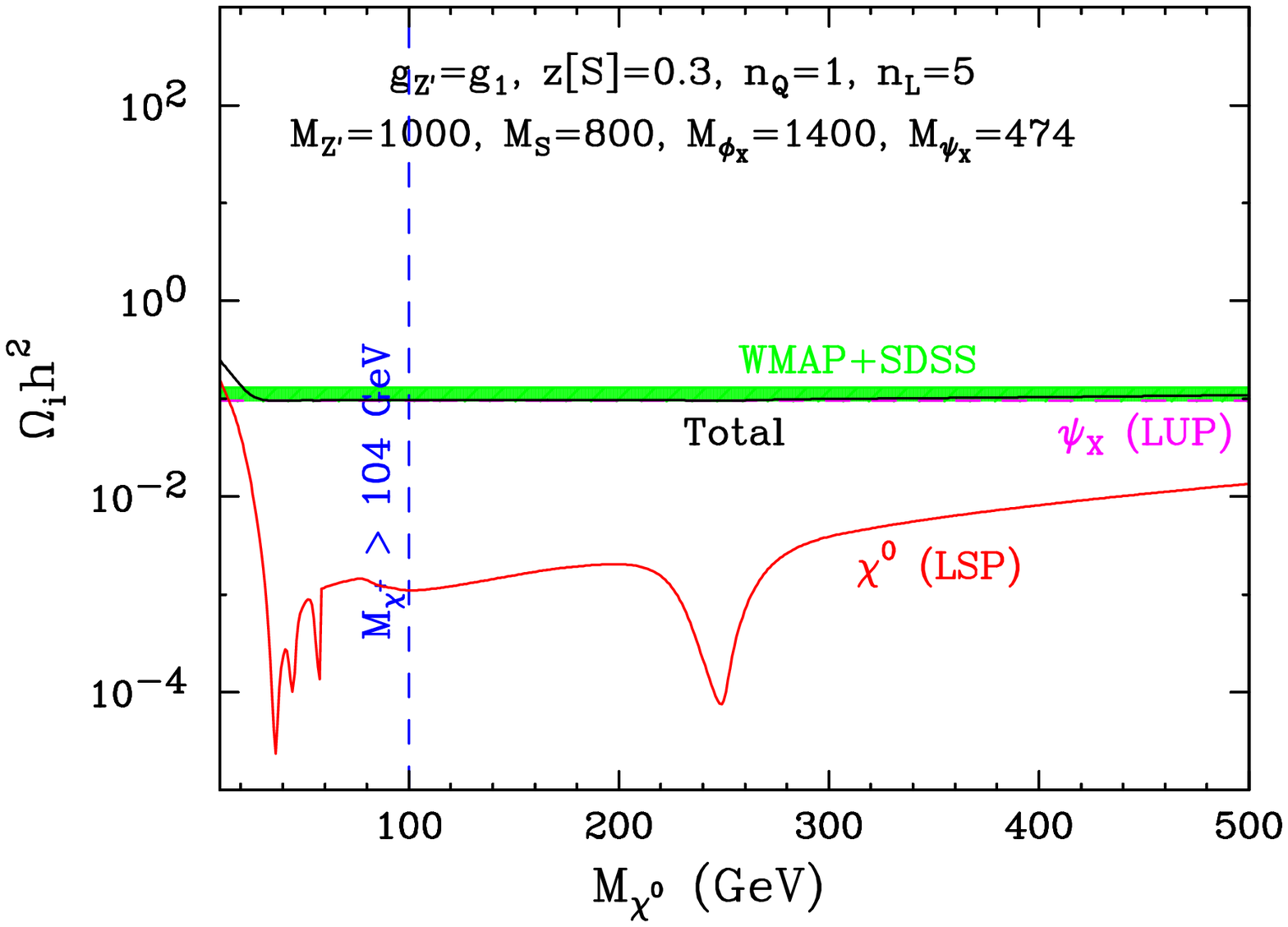}
\includegraphics[width=0.49\textwidth]{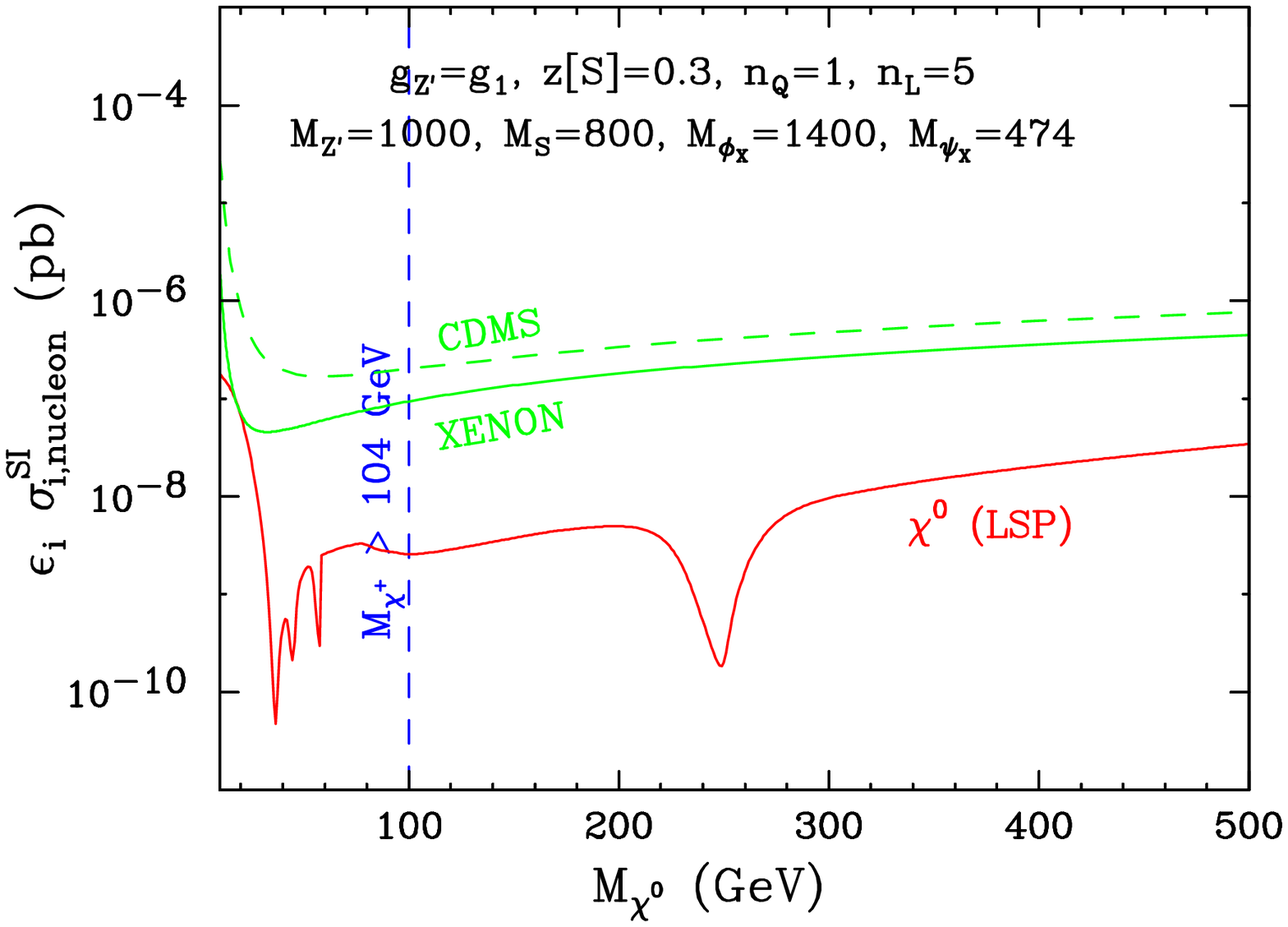}
\caption{(a) Relic density and (b) direct detection effective cross-section ($\epsilon_{\chi^0}\sigma_{\chi^0}$) versus the LSP dark matter ($\chi^0$) mass in the presence of the LUP dark matter $\psix$.
We fix $M_{\psix} = 474 \gev$ and $M_{\phix} = 1400 \gev$, $M_{Z'} = 1000 \gev$.
Dashed curve is the relic density for $\psix$  and solid one is the total relic density of $\psix$ and $\chi^0$.
Vertical line is the exclusion limit by the LEP bound on chargino mass.}
        \label{fig:chi}
\end{figure}

In Figure \ref{fig:chi}, we illustrate the relic density and the effective spin-independent cross-section versus the LSP mass.
We assume the LSP is $\chi^0$ and the LUP is $\psix$ with $\rp$ even and $M_{\psix} = 440 \gev$.
$\phix$ is assumed to be very heavy ($M_{\phix} = 1400 \gev$) so that it is not stable.
The green band is the $2\sigma$ region measured by WMAP+SDSS.
The relic density curve of the $\psix$ is mostly flat since its annihilation cross-section is not sensitive to the $M_{\chi^0}$, and the CDM relic density is dominantly accounted for by the LUP dark matter for the given parameter values.
The poles in the relic density curve of the $\chi^0$ are due to the resonances through $W^\pm$ (in the co-annihilation of $\chi^0$ and $\chi_1^\pm$), $Z$, $h$ and $H/A$.
The $\chi^0$ does not couple to the $Z'$ and we fix $M_{Z'} = 1000 \gev$ for $\psix$ dark matter.
$M_{\chi_1^\pm} \sim M_{\chi^0}$ and $M_{\chi^0} < 100 \gev$ is ruled out by the LEP constraint on chargino mass of $M_{\chi_1^\pm} > 104 \gev$ \cite{LEP2SUSY} as indicated by the vertical line in the figure.
The light Higgs mass also satisfies the LEP bound of $M_{h} > 114 \gev$ over all range of the plot.
We limit the range of the neutralino mass in the plot to be $M_{\chi^0} < 500 \gev$ so that it is always the LSP where slepton masses are fixed at $600 \gev$.
Although the relic density of the LSP dark matter varies with $M_{\chi^0}$, the total relic density is almost constant over the entire $M_{\chi^0}$.
Therefore the finely measured relic density does not give a severe constraint on the LSP property except for its upper bound on the relic density.

Since the $\psix$ does not have a spin-independent cross-section with a nucleus (eq.~(\ref{eq:psixsigma})), the subdominant LSP dark matter is the only {\em detectable} dark matter by the spin-independent nuclear recoil experiment.
The effective cross-section is proportional to the relic density, and the curve is valid for the entire $M_{\chi^0}$ region, since the $\Omega_\text{CDM} h^2 \ne 0.1$ region  ($M_{\chi^0} \lesssim 20 \gev$) is already excluded by the LEP chargino mass bound.
We assumed $\Omega_\text{CDM} h^2 = 0.1$ for calculation of $\epsilon_i = \Omega_i / \Omega_\text{CDM}$ in the effective cross-section ($\epsilon_{\chi^0} \sigma_{\chi^0}$) for the entire range of $M_{\chi^0}$.
The green curves show the current CDMS and XENON experimental constraints on the direct detection.
It is easy to satisfy the constraints since the flux of the detectable dark matter is smaller than that of the entire dark matters.

\begin{figure}[t]
\includegraphics[width=0.49\textwidth]{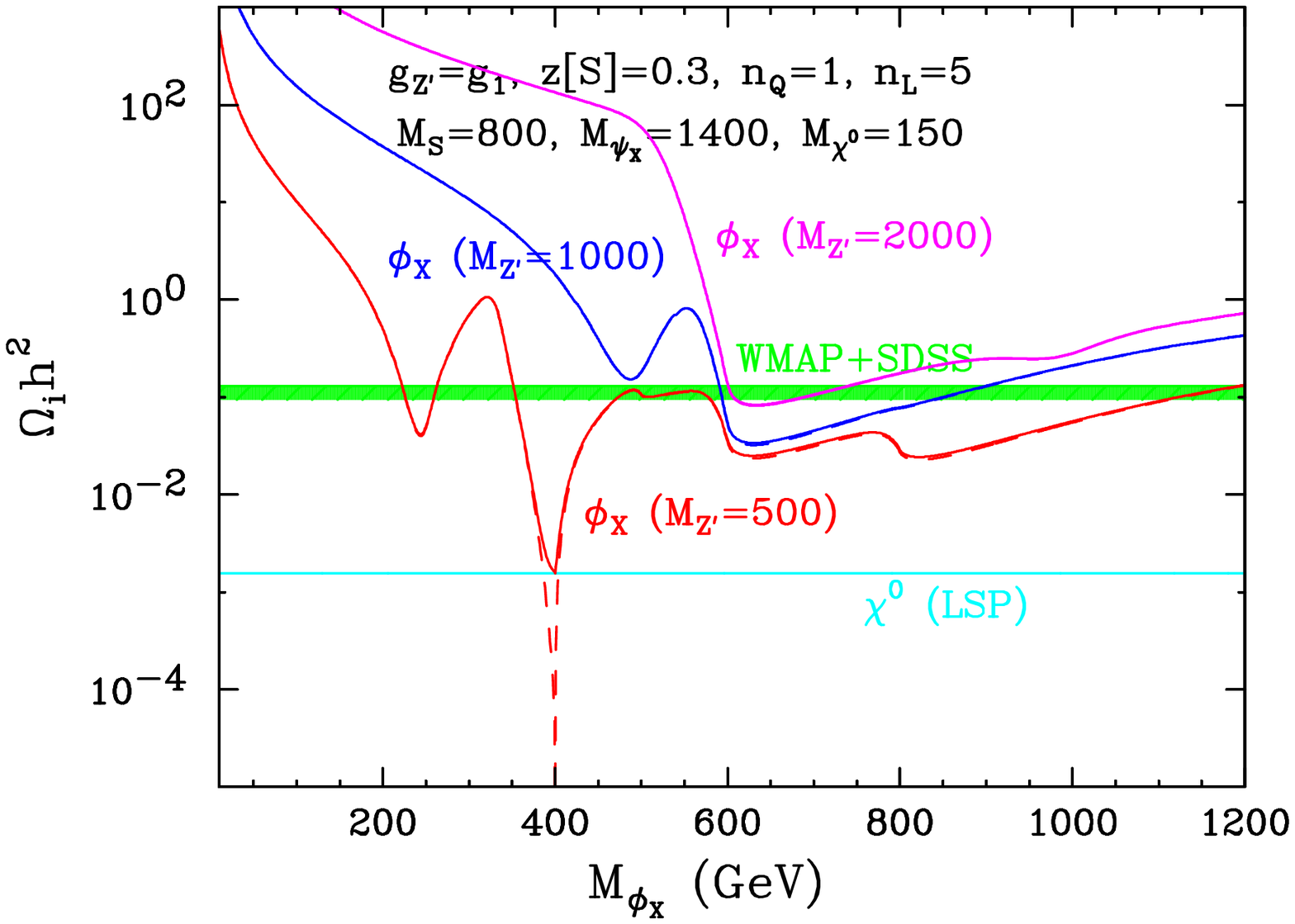}
\includegraphics[width=0.49\textwidth]{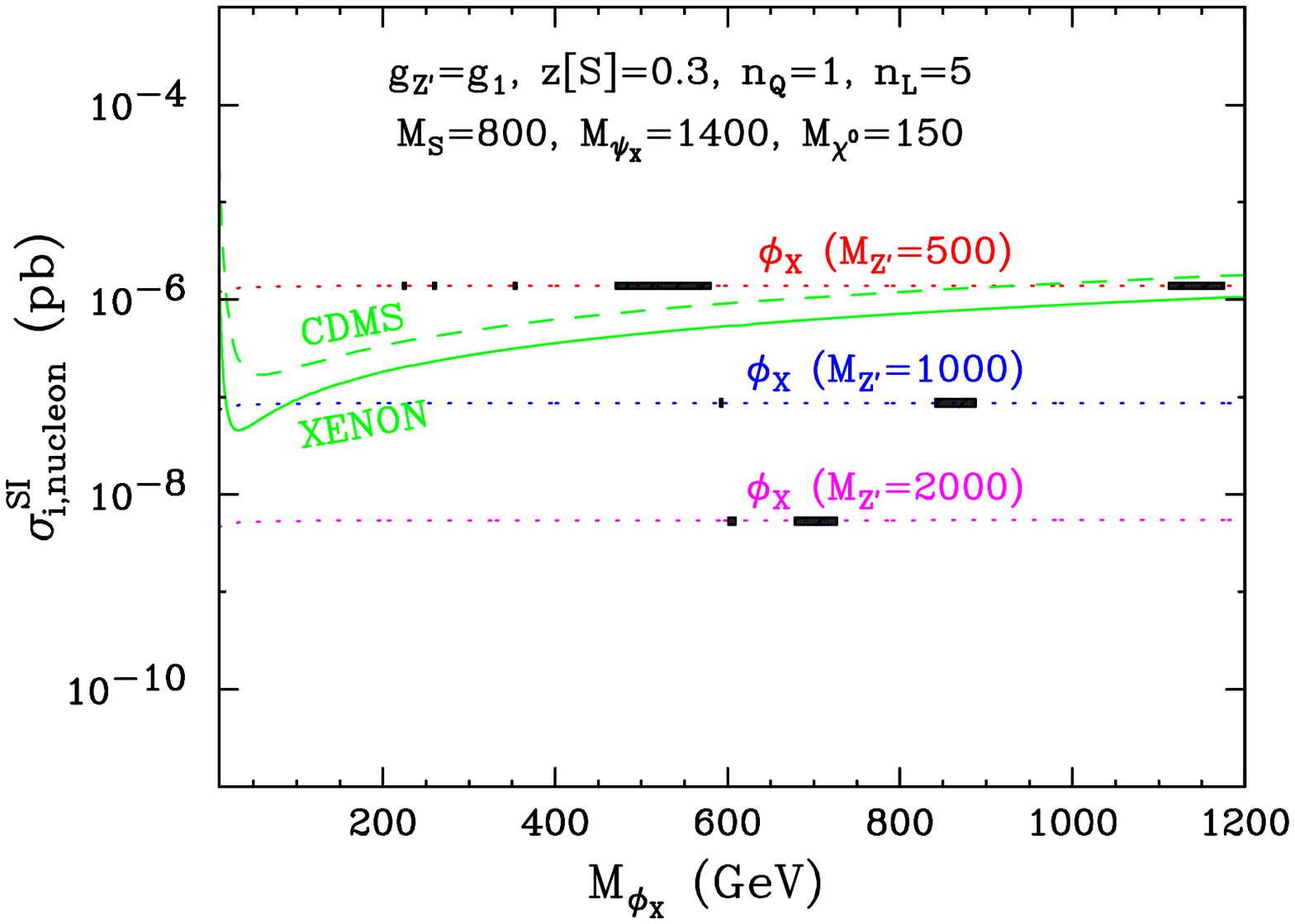}
\caption{(a) Relic density and (b) direct detection cross-section versus the LUP dark matter ($\phix$) mass in the presence of the LSP dark matter $\chi^0$.
We fix $M_{\psix} = 1400 \gev$, $M_{\chi^0} = 150 \gev$, $M_{Z'} = 500, 1000, 2000 \gev$.
Dashed curve is the relic density for $\phix$  and solid one is the total relic density of $\phix$ and $\chi^0$.
Thick black bands are where the total $\Omega_\text{CDM} h^2$ satisfies the WMAP+SDSS limit.}
        \label{fig:phix}
\end{figure}

In Figure \ref{fig:phix}, we illustrate the relic density and the direct detection cross-section versus the LUP mass.
We assume the LUP is $\phix$ with $\rp$ even and the LSP is $\chi^0$.
$\psix$ is assumed to be very heavy ($M_{\psix} = 1400 \gev$) so that it is not stable.
We show curves for $M_{Z'} = 500$, $1000$, $2000 \gev$ to illustrate the dependence on $Z'$ mass.
The relic density curve of the $\chi$ is flat since its annihilation cross-section is not sensitive to the $M_{\phix}$.
The CDM relic density is mostly dominated by the LUP dark matter.

In the $\phix$ curves, we can observe effects of the annihilation channels we discussed in Section \ref{sec:relic}.
We will discuss several of them here.
We see the $Z'$ resonance poles at $M_{\phix} \sim M_{Z'}/2$ as expected.
Among resonances through the $S$, only the $\phix \phix^* \to \chi' \chi'$ for $M_{Z'} = 500 \gev$ is open at $M_{\phix} \sim M_S/2 = 400 \gev$.
This channel is kinematically forbidden for $M_{Z'} = 1000$ or $2000 \gev$, since $M_{\chi'}\sim M_{Z'}\pm M_{1'}/2$ when $M_{Z'} \gg M_{1'}$. 
This $S$ resonance of $M_{Z'} = 500\gev$ is the only region where $\phix$ relic density is smaller than that of $\chi^0$.
The new channel $\phix \phix^* \to \til \ell \til \ell^*$ opens up for $M_{\phix} > M_\text{sleptons}=600 \gev$, making the relic density drop significantly.
This dark matter annihilation channel where the final states are heavy supersymmetric particles (much heavier than the LSP) is a novel feature of this model that can not be found in the single LSP dark matter scenario like the MSSM.
The effect of $\phix \phix^* \to Z'Z'$ is relatively small, but it is still noticeable for $M_{Z'} = 500 \gev$ at $M_{\phix} \sim M_{Z'}$.
The $\phix \phix^* \to S S$ channel for $M_{\phix} > M_S = 800 \gev$ is also distinguishable only for relatively light $Z'$ case.
This is because the dominant contribution in this channel comes from the $F$-term which is proportional to $k^2$.
The $k$ is not an independent variable in our choice of input parameters, and it is small for large $Z'$ mass.
From eqs.~(\ref{eq:B3}) and (\ref{eq:B7}), $k$ is given as $k = g_{Z'} z[S] M_{\psix}/M_{Z'}$.
Due to the variety of channels, it is not difficult to find points with right relic density.
Those points were marked as thick black band in the direct detection plot.

Since the LUP is dominant ($\epsilon_{\phix} \simeq 1$), the effective direct detection curve is practically the cross-section of the LUP, except for the $S$ resonance pole for $M_{Z'} = 500 \gev$ which does not satisfy the relic density constraint anyway.
Therefore, we plot only the direct detection curve for the $\phix$.  
Most part of the $M_{Z'} = 500 \gev$ curve with the choice of parameter values is excluded by the direct detection experiment, while the $M_{Z'} = 1000, 2000 \gev$ curves survive.
The overall size of $\phix$ direct detection rate with TeV scale $Z'$ is comparable to the current ongoing dark matter experiment.
The cross-section decreases if $Z'$ coupling to quarks gets smaller.
If charges are quark-phobic ($z[q_L], z[q_R] \sim 0$), a light $Z'$ would also survive the current constraint by the direct detection and the Tevatron \cite{Abulencia:2006iv} experiment.

Besides the two examples we explored here, there are more possibilities such as all three dark matters coexist as well as only the LUP dark matter exists without the LSP.
Also the LSP and LUP may form the same supersymmetric multiplet (dark matter supermultiplet) in a case, for example, $\psix$ is the LSP and $\phix$ is the LUP.

In the MSSM with a single LSP dark matter, the direct detection and the collider experiment is correlated and it is expected that both experiments detect the dark matter of the same property in the form of nuclear recoil and missing energy, respectively.
With multiple dark matters, the expectations may change.

For the neutralino LSP dark matter, the annihilation cross-section should be larger than the single dark matter scenario since it is responsible for only fraction of the CDM relic density.
While the smaller flux will diminish the chance of the direct detection (in the case $\chi$ is the only detectable dark matter), the increased coupling may enhance the chance of detection at the collider.
The wino or higgsino-like LSP dark matter may also imply interesting phenomenology in the indirect dark matter search.
For example, a large higgsino component could mean large capture rate in the Sun and correspondingly large neutrino fluxes probable in the IceCube experiment \cite{IceCube}.

For the LUP dark matter, the spin-independent direct detection rate by the LUP dark matter itself could be null ($\psix$ case) or large ($\phix$ case).
The LHC/ILC may detect them if they are within the reach of colliders.
$\psix$ and $\phix$ couple only to $Z'$ and $S$, and the prediction will change depending the $Z'$ coupling to quarks and leptons.
For example, the quark-phobic $Z'$ will result in null direct detection rate even for $\phix$ dark matter and the LHC will not be able produce $Z'$ resonance from the typical Drell-Yan process.

Overall, with the presence of multiple dark matters with different sensitivity to the direct detection and to the colliders, the property of the dark matter measured in one experiment may not be consistent with that in the other experiment.

\section{Conclusion}
\label{sec:conclusion}
TeV scale SUSY is a very well motivated new physics scenario.
The minimal version of the supersymmetric SM (MSSM) may not be the correct realization of the TeV scale SUSY though, and its popular dark matter candidate (LSP) may not be the correct or full description of the dark matter.
Various issues of the MSSM actually suggests it needs to be extended to include new ingredients such as new Abelian gauge symmetry $U(1)'$.
The concept of the supersymmetric dark matter may also be extended in the alternative supersymmetric models.

We considered a residual discrete symmetry ($U$-parity) which naturally emerges from the extension of the MSSM with the $U(1)'$.
It provides a new dark matter candidate (LUP) as a stand-alone alternative or coexisting complementary one to the usual LSP dark matter.
We showed that two well-motivated $\ztwo$ symmetries ($R$-parity and $U$-parity) can allow various interesting possibilities and numerically illustrated a few examples that satisfy the experimental constraints for the viable dark matter.

The enriched dark matter properties suggest that the phenomenology for the supersymmetric dark matter (such as physics at the upcoming collider experiments) may be drastically different from the MSSM predictions.
For example, the missing energy at the collider experiment may be originated from two or more massive stable particles, and it will be necessary to develop a technique that can distinguish that from one dark matter case.

\section*{Acknowledgment}
HL and SN are supported by the Department of Energy under grant DE-FG02-97ER41029.
We are grateful to K. Matchev and C. Luhn for useful discussions.
HL is grateful to Professor K. Choi at KAIST, and Professor P. Ko at KIAS for the hospitality in his visits.

\newpage
\appendix

\section{Stable particles}
\label{sec:triangle}

There may be up to three stable massive particles when there are two parities.
Here, we discuss it and categorize the possibilities in our model.
With two $\ztwo$ symmetries ($\rp$ and $\up$), we can classify all component fields into four in terms of $(\rp,\up)$.
\bea
(+,+):~ \text{class}~ A \qquad && (-,+):~ \text{class}~ B \nn \\
(+,-):~ \text{class}~ C \qquad && (-,-):~ \text{class}~ D \nn
\eea
We call the lightest particle of each class as $A_0$, $B_0$, $C_0$, $D_0$, respectively.
($A_0$ is naturally the SM photon.)
The stability of the lightest particles among $\rp$ odd (LSP) and the lightest among $\up$ odd (LUP) are guaranteed by $\ztwo$ symmetries.

The {\it minimal} decay channels of $\ztwo$ odd particles that conserve both parities are
\be
B_0 \to C_0 + D_0, \quad C_0 \to B_0 + D_0, \quad D_0 \to B_0 + C_0.
\ee
A non-minimal decay channel is, for instance,
\be
B_0 \to o_1 C + o_2 D + n_1 A
\ee
where $o_i$ is an odd positive integer and $n_i$ is a non-negative integer.

\begin{figure}[tb]
\includegraphics[width=0.49\textwidth]{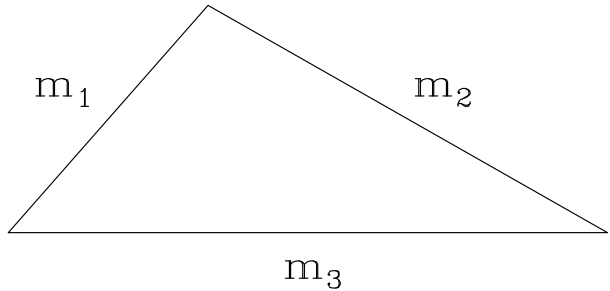}
\caption{Triangle relations of masses ($m_1 + m_2 > m_3$ for $m_1 \le m_2 \le m_3$) to have three stable particles under two $\ztwo$ parities.}
\label{fig:triangle}
\end{figure}

From the above, we can see that
\bea
M_{B_0} < M_{C_0} + M_{D_0} ~\text{guarantees that $B_0$ is stable.} \label{eqn:Bmass} \\
M_{C_0} < M_{B_0} + M_{D_0} ~\text{guarantees that $C_0$ is stable.} \label{eqn:Cmass} \\
M_{D_0} < M_{B_0} + M_{C_0} ~\text{guarantees that $D_0$ is stable.} \label{eqn:Dmass}
\eea
If $M_{B_0}$, $M_{C_0}$, $M_{D_0}$ satisfy a triangle relation (i.e. if all eqs.~(\ref{eqn:Bmass}, \ref{eqn:Cmass}, \ref{eqn:Dmass}) hold as Figure~\ref{fig:triangle}), there will be three stable particles.
Since $\up$ is originated from the gauge symmetry, which commutes with SUSY, any field $F$ and its superpartner $\til F$ share the same $U$-parity.
The relevant fields are then classified as follows.
\begin{enumerate}
\item quarks, leptons, gauge bosons, Higgses: class $A$ $(+,+)$
\item squarks, sleptons, gauginos, higgsinos: class $B$ $(-,+)$
\item $\rp$ even component of $X$: class $C$ $(+,-)$
\item $\rp$ odd component of $X$: class $D$ $(-,-)$
\end{enumerate}
There may be more exotics that belong to $A$ and $B$, or $C$ and $D$ depending on their $U$-parity, but for the sake of simplicity of the discussion we will consider only one exotic superfield $X$.
Then, if triangle relation holds, both $\psix$ and $\phix$ should be two components of the triangle ($C_0$ and $D_0$) and the lightest one among other superparticles ($B_0$) should be the third component of the triangle.
The third component could be the LSP or the Next-to-LSP (NLSP) depending on the relative masses of $B_0$ and $D_0$.

To illustrate stability of three particles with triangle relation, let us assume $M_{\chi} = 500 \gev$, $M_{\psix}=400 \gev$ and  $M_{\phix}=300 \gev$, and assume $\chi$ is the NLSP.
We assign $\rp$ even to $\phix$ and $\rp$ odd to $\psix$.
$\psix$ $(-,-)$ is stable because it is the LSP;
$\phix$ $(+,-)$ is stable because it is the LUP;
$\chi$ $(-,+)$ is stable because its only decay channel $\chi \to \psix \phix$ is closed kinematically.
This corresponds to the Case III of Table~\ref{tab:triangleExamples} which contains the complete list of triangle relation cases.
It is understood that when the triangle relation does not hold, stable massive particles will be less than three.

\begin{table}[bt]
\begin{center}
{\small
\begin{tabular}{|c||c|c|c|c|c|c|}
\hline
& I & II & III & IV & V & VI \\
\hline
$\rp$ odd $X$ ($D_0$) & $m_1$ {\tiny (LUP, LSP)} & $m_1$ {\tiny (LUP, LSP)} & $m_2$ {\tiny (LSP)}      & $m_2$ {\tiny (LUP)} & $m_3$            & $m_3$       \\
$\rp$ even $X$ ($C_0$) & $m_2$            & $m_3$            & $m_1$ {\tiny (LUP)}      & $m_3$       & $m_1$ {\tiny (LUP)}      & $m_2$ {\tiny (LUP)} \\
$\chi$ ($B_0$)         & $m_3$            & $m_2$            & $m_3$            & $m_1$ {\tiny (LSP)} & $m_2$ {\tiny (LSP)}      & $m_1$ {\tiny (LSP)} \\
\hline
\end{tabular}
\caption{Possible triangle relations.
$\chi$ ($B_0$) is the lightest among superparticles possibly except for the $\rp$ odd component of $X$ ($D_0$).
A triangle relation $m_1 + m_2 > m_3$ is assumed ($m_1 \le m_2 \le m_3$) among masses of $\psix$, $\phix$ and $\chi$. 
The LSP and the LUP can be one particle where the other two particles are NLSP and Next-to-LUP (NLUP).
$m_1$ should be the mass of the LSP or the LUP (or both), and $m_3$ can not be that of LSP or LUP.
\label{tab:triangleExamples}}
}
\end{center}
\end{table}

\section{Additional lagrangians}
\label{sec:lagrangians}
Here, we list the lagrangians from the $U(1)'$ symmetry and new field contents.
We omit doublet Higgs and higgsino terms which has vanishing $U(1)'$ charges in our setup.
$f_L$ and $f_R$ represent all the MSSM chiral fields including three Dirac neutrinos.

\begin{enumerate}
\item Fermion-Fermion-$Z'$
\bea
{\cal L}_1 &=& - \frac{1}{2}g_{Z'} z[f_L] Z'^\mu \overline f \gamma_\mu (1-\gamma^5) f  - \frac{1}{2}g_{Z'} z[f_R] Z'^\mu \overline f \gamma_\mu (1+\gamma^5) f \nn \\ 
&& + \frac{1}{2}g_{Z'} z[X]Z'^\mu \overline {\psix} \gamma_\mu  \gamma^5  \psix 
+ \frac{1}{2}g_{Z'} z[S]Z'^\mu \overline {\til S} \gamma_\mu \gamma^5 \til S   
\eea

\item Scalar-Scalar-$Z'$
\bea
{\cal L}_2 &=& - i g_{Z'} z[f_L] Z'^\mu ( {\til f}_L^* \partial_\mu {\til f}_L - \partial_\mu {\til f}_L^* {\til f}_L )
- i g_{Z'} z[f_R] Z'^\mu ( {\til f}_R^* \partial_\mu {\til f}_R - \partial_\mu {\til f}_R^* {\til f}_R ) \nn\\ 
&&- i g_{Z'} z[X] Z'^\mu ( \phix^* \partial_\mu \phix - \partial_\mu \phix^* \phix )  
\eea

\item Scalar-Scalar-$Z'$-$Z'$
\bea
{\cal L}_3 &=& g_{Z'}^2 z[f_L]^2 Z'^\mu Z'_\mu {\til f}_L^* {\til f}_L + g_{Z'}^2 z[f_R]^2 Z'^\mu Z'_\mu {\til f}_R^* {\til f}_R \nn\\ 
&& + g_{Z'}^2 z[X]^2 Z'^\mu Z'_\mu \phix^*  \phix 
+ g_{Z'}^2 z[S]^2 Z'^\mu Z'_\mu S^* S 
\label{eq:B3}
\eea

\item Scalar-Fermion-$\til Z'$
\bea
{\cal L}_4 &=& -\frac{\sqrt{2}}{2} g_{Z'} z[f_L] \left( \overline {\til Z'} ( 1-\gamma^5 ) f \til{f}_L^*  
                                + \overline f ( 1+\gamma^5 ) \til Z' \til{f}_L \right) \nn \\
&&+ \frac{\sqrt{2}}{2} g_{Z'} z[f_R] \left( \overline {\til Z'} ( 1+\gamma^5 ) f \til{f}_R^*  
                                + \overline f ( 1-\gamma^5 ) \til Z' \til{f}_R \right) \nn \\
&&-\frac{\sqrt{2}}{2} g_{Z'} z[X] \left( \overline {\til Z'} ( 1-\gamma^5 ) \psix \phix^*  
                                + \overline {\psix} ( 1+\gamma^5 ) \til Z' \phix \right) \nn \\
&&-\frac{\sqrt{2}}{2} g_{Z'} z[S] \left( \overline {\til Z'} ( 1-\gamma^5 ) \til S S^*  
                                + \overline {\til S} ( 1+\gamma^5 ) \til Z' S \right)
\eea

\item $D$-term
\bea
{\cal L}_5 &=& -\frac{1}{2} g_{Z'}^2 \left( z[f_L]{\til f}_L^* {\til f}_L 
                        - z[f_R]{\til f}_R^* {\til f}_R
                         + z[X] \phix^* \phix + z[S] S^* S \right)^2 
\eea

\item $F$-term
\bea
{\cal L}_6 &=& - k^2 S^* S \phix^* \phix
- \frac{k^2}{4} (\phix^* \phix)^2    
\eea

\item Yukawa
\bea
{\cal L}_7 &=& -\frac{k}{2}  \left( \overline {\til S} ( 1-\gamma^5 ) \psix \phix  
                                + \overline{\psix} ( 1+\gamma^5 ) \til{S} \phix^* \right) \nn \\
                                && - \frac{k}{4} \left( \overline \psix (1-\gamma_5) \psix S + \overline \psix (1+\gamma_5) \psix S^* \right)
\label{eq:B7}
\eea

\item Soft term
\bea
{\cal L}_8 &=& \left( - \frac{1}{2} M_{1'} \til Z' \til Z' -A_{SXX} S \phix \phix + h.c. \right) - m_S^2 S^* S - m_{\phix}^2 \phix^* \phix
\eea

\end{enumerate}

\newpage

\end{document}